\documentclass{article}

\usepackage{arxiv}

\usepackage[utf8]{inputenc} 
\usepackage[T1]{fontenc}    
\usepackage{hyperref}       
\usepackage{url}            
\usepackage{booktabs}       
\usepackage{amsfonts}       
\usepackage{microtype}      
\usepackage{lipsum}		
\usepackage{graphicx}
\usepackage{natbib}
\usepackage{doi}
\usepackage{caption}
\usepackage{subcaption}

\usepackage{lineno}     
\usepackage{mathrsfs}
\usepackage{amsthm,amsmath,amsfonts,amssymb}
\usepackage{placeins}
\newcommand{\beq}{\begin{equation}}
\newcommand{\eeq}{\end{equation}}
\newcommand{\beqa}{\begin{eqnarray}}
\newcommand{\eeqa}{\end{eqnarray}}

\title{ Bayesian Estimation of Extreme Quantiles and the Exceedance Distribution for Paretian Tails} 
\author{ 
\hspace{1mm}Douglas E. Johnston\\
	Department of Applied Mathematics\\
	Farmingdale State College\\
	Farmingdale, NY 11735 \\
	\texttt{douglas.johnston@farmingdale.edu} \\
}

\date{}   

\renewcommand{\headeright}{}     
\renewcommand{\undertitle}{}     

\begin{document}
\maketitle

\begin{abstract}
Estimating extreme quantiles is an important task in many applications, including financial risk management and climatology. More important than estimating the quantile itself is to insure zero coverage error, which implies the quantile estimate should, on average, reflect the desired probability of exceedance. In this research, we show that for unconditional distributions isomorphic to the exponential, a Bayesian quantile estimate results in zero coverage error. This compares to the traditional maximum likelihood method, where the coverage error can be significant under small sample sizes even though the quantile estimate is unbiased. More generally, we prove a sufficient condition for an unbiased quantile estimator to result in coverage error.  Interestingly, our results hold by virtue of using a Jeffreys prior for the unknown parameters and is independent of the true prior.  We also derive an expression for the distribution, and moments, of future exceedances which is vital for risk assessment. We extend our results to the conditional tail of distributions with asymptotic Paretian tails and, in particular, those in the Fréchet maximum domain of attraction. We illustrate our results using simulations for a variety of light and heavy-tailed distributions.
\end{abstract}


\section{Introduction}
One of the key statistical tasks in many applications is the estimation of prediction intervals, or quantiles. In fields as diverse as target detection (\citet{broadwater2010}), communication systems (\citet{ResRootz2000}), image analysis (\citet{SJRoberts2000}), power systems (\citet{shenoy2015}), and population studies (\citet{Anderson3252}), estimating extreme quantiles is important.  In the field of finance, {\em VaR} analysis is a standard tool used by risk-managers and bank supervisors  (\citet{dej_pmd_SPM2011}).  In hydrology, N-year return level studies are the norm (\citet{Nerantzaki_2022}). Both of these require quantile estimates, for setting proper risk constraints, but the distribution of exceedances above the quantile estimate is a vital risk measure too. 

Typically, the underlying data exhibits extreme, or heavy-tailed, behavior where the underlying probability distribution decays according to a power-law.  The exponent, or tail-index, is usually estimated from past data using methods such as maximum-likelihood or the method of moments (\citet{smith1987}, \citet{hosking1987}) and analyses of those quantile estimators have been studied (\citet{buishand1989}). Once an estimate for the tail-index is obtained, it is typically used to make inferences regarding prediction intervals, or quantile exceedances, of future events although poor small sample size performance has been noted (\citet{coles_1999}).  While Bayesian methods have been employed, the focus has primarily been on the adoption of prior information and the use of loss functions (\citet{coles_1996}) or in the context of quantile regression (\citet{yu2001}). Estimating the distribution of exceedances over an order statistic, or a record, has also been studied (\citet{gumbel1950}, \citet{wesolowski1998}) where distribution-free methods have been derived. 

In terms of our research, there are two related problems that we wish to solve.  One is to use all the data and estimate a quantile, and exceedance distribution, from an unconditional distribution. This has the benefit of potentially eliminating small sample size problems at the expense of a very high quantile. For example, if an annual 99\% quantile is needed (e.g., a 1 in 100 year event forecast), and there are 100 samples per year, then the 99.99\% quantile of the unconditional distribution is required.  In this case our results are limited to distributions that are isomorphic to the exponential distribution.  

An alternative to using all of the data and an unconditional model, is to focus solely on the conditional tail of the distribution which, as we will show, can be asymptotically modeled using the exponential distribution.  In particular, this opens the door to unconditional distributions that exhibit power-law tail behavior, such as heavy-tailed distributions in the Fréchet maximum domain of attraction, at the cost of restricting our samples to extreme events. Our paper proceeds as follows.

In Section \ref{Sec2.1}, we drive a Bayesian quantile estimate by marginalizing the predictive distribution of future samples over the parameter posterior given past observations. In particular, we show that for unconditional distributions isomorphic to the exponential, a Bayesian quantile estimate results in zero coverage error (ZCE), which means that a quantile estimate at the $\alpha$-percentile will be exceeded, on average $1-\alpha$\% of the time. This compares to the traditional maximum likelihood method (MLM), where the coverage error can be significant under small sample sizes even though the maximum likelihood estimate is unbiased.

A similar quantile estimate was obtained in (\citet{YuAlly2009}), although they derived their estimator by requiring zero-coverage error and then solving for the appropriate estimator.  Our result holds by virtue of using a Bayeisan approach with the Jeffreys prior for the unknown parameters and it is independent of the true prior for the model parameters.  This illustrates the probability matching criteria of the Jeffreys prior in computing Bayesian prediction intervals (\citet{sweeting2000}). In addition, we prove a sufficient condition for an unbiased quantile estimator to result in coverage error. Given that most distributions encountered in practice satisfy the condition, this implies that seeking an unbiased estimator is not ideal.  

In Section \ref{Sec2.2}, we derive an expression for the distribution, and moments, of future exceedances. We obtain a new discrete distribution, which we term $BEG$ and it resembles the distribution for record exceedances found in (\citet{wesolowski1998}, \citet{bairamov1996}) although in our case we deal with the exceedance over a quantile estimate, which is not necessarily related to past records.  We compare to the distribution obtained with the MLM and that obtained in (\citet{gumbel1950}), which is for exceedances over an order statistic. The latter is a distribution-free method but only readily applies for levels of $\alpha$ associated with empirical quantiles. As shown in (\citet{hallriect2001}), such non-parametric methods can be improved on, to allow extrapolation to extreme quantiles, but they do not result in ZCE. Our method not only results in ZCE but has lower moments and variance. In Section \ref{Sec2.3}, we show that our results can be extended to any distribution that can be transformed into the exponential, which includes for example, the Rayleigh and standard Pareto distributions.

In Section \ref{Sec3}, we apply our results to the conditional tail of distributions that are in the Fréchet maximum domain of attraction, which includes many heavy-tailed distributions.  We model the point process of exceedances, over a threshold $u$, as a Poisson process where the predictive posterior distribution for the number of future exceedances is a negative binomial and the exceedance level is a Lomax (Pareto 2) distribution.  From, this we derive the Bayesian estimate with zero coverage error. Lastly, in Section \ref{Sec4}, we illustrate the performance for a variety of distributions, both heavy-tailed and not.

\section{Unconditional Quantile Estimation}
\label{Sec2}

\subsection{Bayesian and ML Quantile Estimation}
\label{Sec2.1}

Let $X_1, \cdots, X_n \equiv X_{1:n}$ and $Y_1, \cdots, Y_N \equiv Y_{1:N}$ be independent random variables with common distribution function $F(\tau/\lambda) = 1 - e^{-\lambda \tau}$. From the past observations, $x_{1:n}$, we estimate quantiles of future observations, $y_{1:N}$, based on a predictive cdf $\hat{F}( y / x_{1:n} )$ that is inverted at the level $\alpha \in (0,1)$ so that $\hat{\eta}_\alpha = \hat{F}^{-1}(\alpha/  x_{1:n})$.  One traditional method to compute a predictive cdf is the maximum likelihood (ML) method (MLM) which uses the underlying cdf of the training data, $F(x/\lambda)$, where the ML estimates for unknown parameters are substituted. For the exponential case, this results in  
\begin{equation}
    {\hat F}( y / x_{1:n} ) = 1 - e^{- \hat{\lambda} y}, \;\;\;\;  \hat{\lambda} = \frac{n}{\sum x_i},
\end{equation}
where the summation, $\sum x_i$, runs from $i=1$ to $n$. The ML quantile estimate is then derived as
\begin{equation}
    \hat{\eta}_{\alpha,ML} = - \frac{ \log( 1 - \alpha)}{\hat{\lambda}} = -\frac{\log(1-\alpha)}{n}  \sum x_i.
\end{equation}
and the MLM estimate is unbiased,
\begin{equation}
    E\{ \hat{\eta}_{\alpha,ML} \} = -\frac{\log(1-\alpha)}{n} \sum E\{x_i\}  = - \frac{ \log( 1 - \alpha)}{\lambda} = \eta_\alpha,
\end{equation}
and $\hat{\eta}_{\alpha,ML}$ converges to $ \eta_\alpha$ in probability.

A Bayesian approach is to compute the expected cdf under the posterior for $\lambda$, $\mathcal{P}( \lambda  / x_{1:n})$, which is then inverted to obtain quantile estimates.  For $n$ i.i.d. samples from Exp$(\lambda)$ and Jeffreys prior ($\mathcal{P}(\lambda) \propto 1/\lambda$), the posterior is a gamma distribution, $\Gamma( n, \sum x_i )$ and the predictive cdf is then defined as
\begin{equation}
    \hat{F}( y / x_{1:n} ) = E_{\lambda/x_{1:n}} [F( y / \lambda ) ] = \int F( y / \lambda )  \mathcal{P} ( \lambda  / x_{1:n} ) d\lambda
\end{equation}
which is the expected cdf marginalized over the posterior. Substituting in the distributions we have
\begin{equation}
    \hat{F}( y / x_{1:n} ) = \frac{ (\sum x_i)^n }{\Gamma(n)} \int_0^\infty ( 1 - e^{-\lambda x})  \lambda^{n-1} e^{- \lambda \sum x_i} d\lambda
\end{equation}
and the predictive posterior is 
\begin{equation}
\label{eq:bayes_cdf}
    \hat{F}( y / x_{1:n} ) = 1 - \frac{1}{ (1 + \frac{y}{\sum x_i})^n }.
\end{equation}
This is a Pareto II (Lomax) distribution which we invert to obtain
\begin{equation}
    \hat{\eta}_{\alpha,Bayes} = \left( \frac{1}{(1-\alpha)^{1/n}} - 1 \right) \sum x_i
\end{equation}
and we can compare to the unbiased ML
\begin{equation}
    \hat{\eta}_{\alpha,ML} = \log\left(\frac{1}{(1-\alpha)^{1/n}}\right)  \sum x_i.
\end{equation}
Both estimators can be written as $\Psi \sum x_i$, with $\Psi > 0$, and, since $\log(x) < x - 1$ for $x \ne 1$, $\hat{\eta}_{\alpha,Bayes} > \hat{\eta}_{\alpha,ML}$ with $0 < \alpha < 1$ and, therefore, the Bayesian quantile estimate is biased.  That said,  $\hat{\eta}_{\alpha,Bayes} \to \hat{\eta}_{\alpha,ML}$ as $n \to \infty$ so it is asymptotically unbiased and converges in probability to the true quantile.

The convergence of the quantile estimators implies that the conditional probability of a future sample, $Y$, exceeding $\hat{\eta}_{\alpha}$ is $e^{-\lambda \hat{\eta}_{\alpha}}$ and, since $\hat{\eta}_{\alpha} \to \eta_{\alpha} = -\frac{\log(1-\alpha)}{\lambda}$ as $n \to \infty$, we have $P[ Y > \hat{\eta}_{\alpha} / x_{1:n} ] \to 1 - \alpha$. By the law of iterated expectations, we can extend this to the unconditional probability of an exceedance.  Given enough samples, either quantile estimate can be reliably used; however, for small or even modest sample sizes and high quantiles, the quantile estimates, and their coverage error, can be substantially different. 

Given a quantile estimate that is a function of our data, $\hat{\eta}_{\alpha}(\sum x_i)$, we wish to compute $E[ N_\alpha(y_{1:N}) ]$, the expected number of test samples that are greater than $\hat{\eta}_{\alpha}(\sum x_i)$. For convenience, we drop the explicit dependence of $\hat{\eta}_{\alpha}$ and $N_\alpha$ on the samples $x_{1:n}$ and $y_{1:N}$, respectively, and use $\Sigma \equiv \sum_i^n x_i$. The expectation is with respect to the training and test samples as well as the model parameter,
\begin{equation}
    E[ N_\alpha ] = E_{\lambda, \Sigma, y_{1:N}} [ N_\alpha ] = E_{\lambda, \Sigma} \left[ E_{y_{1:N}/\lambda, \Sigma} [ N_\alpha]\right].
\end{equation}

Conditioned on $\lambda, \Sigma$, the cdf of a test sample is $F(y /\lambda, \Sigma)$ and $N_\alpha$ is a binomial random variable with 
\begin{equation}
E_{y_{1:N}/\lambda, \Sigma} [ N_\alpha] = N (1 - F(\hat{\eta}_{\alpha} /\lambda, \sigma) ) = N \mathcal{P}(Y > \hat{\eta}_\alpha / \lambda, \Sigma )    
\end{equation}
and therefore
\begin{equation}
 E[ N_\alpha ] = N \int_{\lambda, \Sigma }  (1 - F(\hat{\eta}_{\alpha} /\lambda, \Sigma ) f( \lambda, \Sigma) d\lambda d\Sigma  = N \mathcal{P}(Y > \hat{\eta}_\alpha)    
\end{equation}
where $f( \lambda, \Sigma)$ is the joint density of the training samples and model parameters. We integrate over $\Sigma$ first using $f(\lambda, \Sigma) = f( \Sigma/ \lambda) f(\lambda)$ where $f(\lambda)$ is the true, unknown prior distribution for $\lambda$ not the prior used to form the quantile estimate.  Therefore,
\begin{equation}
\label{eq:freq}
 E[ N_\alpha ] = N \int_{\lambda} \left[ \int_{\Sigma}  (1 - F(\hat{\eta}_{\alpha} /\lambda, \Sigma ) f( \Sigma/ \lambda)  d\Sigma \right] f(\lambda) d\lambda      
\end{equation}
 and, since both quantile estimators are of the form $\Psi \Sigma$, we can write 
\begin{equation}
E_{y_{1:N}/x_{1:n}, \lambda} [ N_\alpha]  = N e^{-\lambda \Psi \Sigma}.
\end{equation}

We first marginalize over $\Sigma$ conditioned on $\lambda$ which has a gamma distribution, $f(\Sigma / \lambda) \sim \Gamma(n, \lambda)$, to solve the inner integral of Equation \ref{eq:freq} as
\begin{equation}
\int_0^\infty e^{-\lambda \Psi \Sigma} \frac{\lambda^n}{\Gamma(n)} (\Sigma)^{n-1} e^{-\lambda \Sigma} d\Sigma = \frac{1}{(\Psi + 1)^n}.
\end{equation}
Interestingly, this integral does not involve $\lambda$, which we have yet to marginalize over and, therefore, regardless of $f(\lambda)$, we can state 

\beq
E[ N_\alpha ] = \frac{N}{(\Psi + 1)^n}.
\eeq
Further, since the Bayesian quantile estimator has $\Psi = \frac{1}{(1-\alpha)^{1/n}}-1$ we would obtain 
\begin{equation}
    E[ N_\alpha ] = N (1-\alpha) = N \mathcal{P}( y > \hat{\eta}_{\alpha, Bayes} )
\end{equation}
which shows that the Bayesian quantile estimator has zero coverage error (ZCE) with $\mathcal{P}( y > \hat{\eta}_{\alpha, Bayes}) = 1 - \alpha$, for all $n$.  It is worth noting that achieving zero coverage error is achieved by virtue of using the Jeffreys prior in forming the Bayesian estimate. The ML quantile estimator, while unbiased, results in $E[ N_\alpha ] = N \mathcal{P}( y > \hat{\eta}_{\alpha, ML} ) > N(1-\alpha)$. That is, under repeated sampling, the ML quantile estimate will produce too many exceedances.

\begin{figure}[ht]
	\centering
    \includegraphics[width=.715\textwidth]{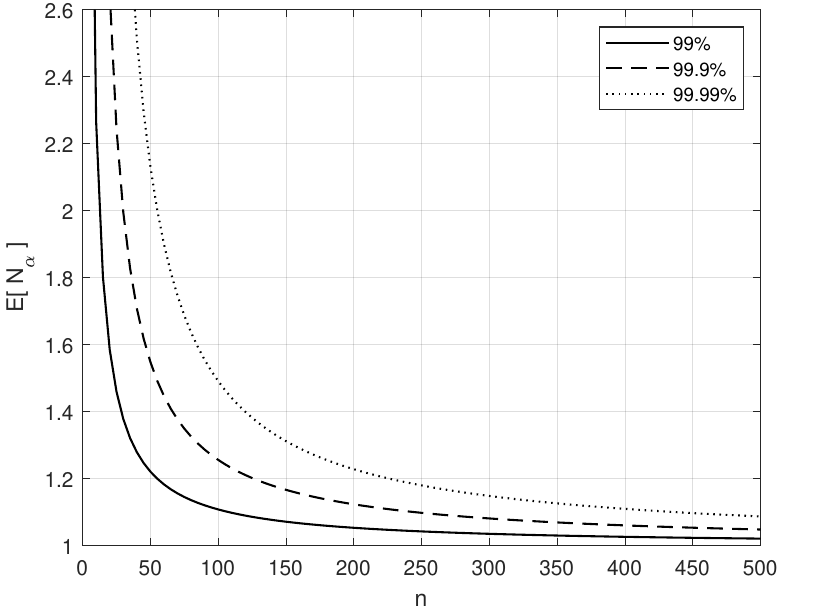}
    \caption{ $E[ N_\alpha ]$ vs $n$ for MLM with $N=1/(1-\alpha)$ and $\alpha = 99\%$, $99.9\%, 99.99\%$}	\label{fig:fig1}
\end{figure}

Figure \ref{fig:fig1} shows the expected exceedences using the MLM quantile estimator versus the number of training samples, $n$, for $\alpha = 99\%$, $99.9\%$, and 99.99\% where the number of test samples was chosen to be $N = 1/(1-\alpha)$.  In this case, the Bayes estimator will have $E[ N_\alpha ] = 1$ so we may justifiably call the Bayesian quantile estimate an N-year return level. We can see that the MLM's probability coverage error persists and that a large number of training samples ($n \geq 250$) are needed for effective high-quantile estimation.  

A natural question to ask is under what conditions will $\mathcal{P}( y > \hat{\eta}_{\alpha} ) > (1-\alpha)$. A sufficient condition is if the quantile estimator is unbiased and the cdf $F(.)$ is analytic and concave. Since the distribution tails we typically encounter have either exponential or power-law decay, such as the Pareto, the assumptions on $F$ are fairly benign.

Since $N_\alpha$ is a binomial RV, $E[N_\alpha / \hat{\eta}_\alpha)] = N (1 - F(\hat{\eta}_\alpha))$.  If we assume that $F$ is analytic and concave then
\begin{equation}
    F(\hat{\eta}_\alpha) < F({\eta}_\alpha) + F'({\eta}_\alpha)(\hat{\eta}_\alpha-{\eta}_\alpha) \;\;\;\; \forall \; \hat{\eta}_\alpha \ne {\eta}_\alpha
\end{equation}
and
\begin{equation}
    F(\hat{\eta}_\alpha) = F({\eta}_\alpha) + F'({\eta}_\alpha)(\hat{\eta}_\alpha-{\eta}_\alpha) + HOT
\end{equation}
with $HOT < 0$. Given
\begin{equation}
    E[N_\alpha] = E_{\hat{\eta}_\alpha}[ E[N_\alpha/\hat{\eta}_\alpha]] = N  E_{\hat{\eta}_\alpha}[1 - F(\hat{\eta}_\alpha)],
\end{equation}
we have
\begin{equation}
    E[N_\alpha] = N E_{\hat{\eta}_\alpha}[ 1 - (F({\eta}_\alpha) + F'({\eta}_\alpha)(\hat{\eta}_\alpha-{\eta}_\alpha) + HOT) ].
\end{equation}
Since the estimate is assumed unbiased, this results in
\begin{equation}
    E[N_\alpha] = N ( 1 - F({\eta}_\alpha) - E[HOT] ).
\end{equation}
Since the $HOT < 0 \; \forall \; \hat{\eta}_\alpha \ne {\eta}_\alpha$ then as long as $\mathcal{P}(\hat{\eta}_\alpha \ne {\eta}_\alpha) > 0$, $E[HOT] < 0$ and
\begin{equation}
    E[N_\alpha] > N ( 1 - F({\eta}_\alpha)) = N(1-\alpha).
\end{equation}
Therefore, under mild conditions, an unbiased quantile estimator results in probability coverage error.

\subsection{The Distribution of Exceedances}
\label{Sec2.2}

Similar to computing $E[N_\alpha]$, we can calculate $\mathcal{P}(N_\alpha = k)$ for $k=0,1,...N$. Conditioned on $\Sigma$ and $\lambda$, $N_\alpha$ is a Binomial RV $\sim Bin(k; N, p$) with $p = \mathcal{P}( Y > \hat{\eta}_{\alpha} ) = e^{-\lambda \Psi \Sigma}$.  Thus, 
\begin{equation}
\mathcal{P}(N_\alpha = k / \Sigma, \lambda, N) = {N \choose k} (e^{-\lambda \Psi \Sigma})^k ( 1- e^{-\lambda \Psi \Sigma})^{N-k}
\end{equation}
which, as before, we integrate over the joint density $f( \Sigma, \lambda) = f( \Sigma / \lambda) f(\lambda)$. 
First integrating over $f( \Sigma / \lambda)$ yields
 \begin{equation}
     \int_0^\infty {N \choose k} (e^{-\lambda \Psi \Sigma})^k ( 1- e^{-\lambda \Psi \Sigma})^{N-k} \frac{\lambda^n}{\Gamma(n)} (\Sigma)^{n-1} e^{-\lambda \Sigma} d\Sigma
 \end{equation}
and using the binomial expansion produces an expression independent of $\lambda$ and therefore
\begin{equation}
    \label{beg_dist}
    \mathcal{P}(N_\alpha = k ) = {N \choose k} \sum_{j=0}^{N-k} (-1)^{N-k-j} {N-k \choose j} \frac{1}{(\Psi (N-j) + 1)^n}.
\end{equation}
We term this the Binomial-Exponential-Gamma (BEG) distribution, BEG$(k; n, N, \alpha)$, where $\Psi$ depends on the quantile estimation method.  The $k$'th moment for the BEG distribution can be derived as
\begin{equation}
    E[ N_\alpha^k ] = \sum_{i=0}^k {k \brace i} N^{\underline{i}} (i \Psi + 1 )^{-n}
\end{equation}
where ${k \brace i}$ are Stirling numbers of the second kind and $N^{\underline{i}}$ is the $i$'th falling power of $N$. We note that all Bayes moments are lower than with the MLM. Also, given the true binomial distribution of exceedances has moments
\begin{equation}
    E[ N_\alpha^k ] = \sum_{i=0}^k {k \brace i} N^{\underline{i}} (1-\alpha)^{i},
\end{equation}
the Bayesian choice for $\Psi$ can be viewed through the lens of the method of moments, where we match the mean of the exceedance distribution. We can derive the variance of the number of exceedences as
\begin{equation}
    Var[ N_\alpha ] =  E[ N_\alpha ] (1 -  E[ N_\alpha ]) +\frac{N(N-1)}{(2\Psi+1)^n}.
\end{equation}

\begin{figure}[ht]
	\centering
    \includegraphics[width=1\textwidth]{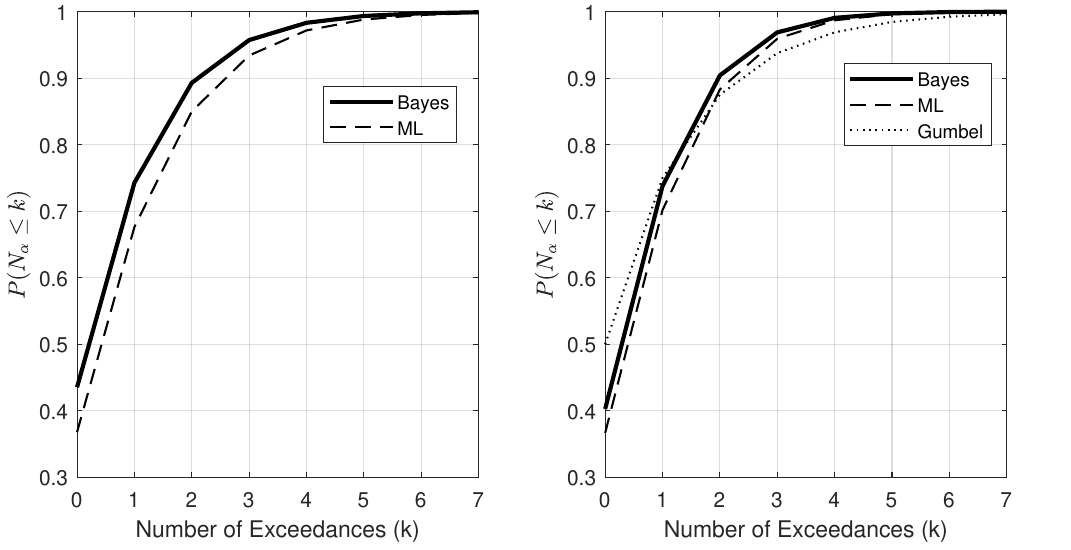}
    \caption{ $BEG(k; 50, 100, 99\%)$, left, and $BEG(k; 100, 100, 99\%)$, right, along with GvS: $w(100,1,100,k)$ exceedance distributions}	\label{fig:fig2}
\end{figure}

On the left of Figure \ref{fig:fig2} are the theoretical distributions, $BEG(k; 50, 100, 99\%)$, for both the Bayes and ML quantile estimators with $n = 50$ training samples, $N = 100$ test samples and $\alpha = 99\%$.  For all $k$, the $\mathcal{P}(N_\alpha > k )$ is higher for the MLM distribution with $E[N_\alpha] = 1.22$ and $Var[N_\alpha] = 2.11$ compared to the Bayesian distribution with $E[N_\alpha] = 1$ and $Var[N_\alpha] = 1.46$. We note that for this case, since $n < N$ and $\alpha = 1 - 1/N$, an empirical quantile estimate using the highest-order statistic is not directly applicable.

On the right of Figure \ref{fig:fig2} are the $BEG(k; 100, 100, 99\%)$ distributions along with the GvS distribution, $w(100,1,100,k)$, of rare exceedances (\citet{gumbel1950}), which is a distribution-free estimate using, in this case, the first-order statistic for the $\alpha$-quantile since $n=N=1/(1-\alpha)$. While the GvS distribution, like the Bayesian, has mean 1, it's variance is 2 compared to the Bayesian distribution which has variance 1.21.  The GvS distribution has higher probabilities for larger number of exceedences, which is not totally surprising given it is a distribution-free, empirical, quantile estimator. The MLM distribution results in $E[N_\alpha] = 1.11$ and $Var[N_\alpha] = 1.48$.

\subsection{Extending the Exponential Distribution}
\label{Sec2.3}

In general, if $Z$ is from a distribution that can be transformed by a one-to-one transformation to $X = h(Z) \sim Exp(\lambda)$ then we can produce quantile estimates from the transformed data and $\hat{\eta}_\alpha^Z = h^{-1}( \Psi \sum h(z_i) )$.  Because $h$ is one-to-one, $\hat{\eta}_\alpha^Z = h^{-1}( \hat{\eta}_\alpha^X )$ or  $\mathcal{P}( Z > \hat{\eta}_\alpha^Z ) = \mathcal{P}( X > \hat{\eta}_\alpha^X )$. In addition, since $Z > \hat{\eta}_\alpha^Z$ iff $X > \hat{\eta}_\alpha^X$, we have $N_\alpha^X = N_\alpha^Z$ and, therefore, $E[N_\alpha^X] = E[N_\alpha^Z]$.  Therefore, our results apply to any distribution that is isomorphic to the exponential distribution and the distribution of exceedances will be invariant.  

The MLM estimate of the quantile, itself, is invariant under the 1-1 RV transformation as long as the transformation does not depend on the parameter.  Given $f(x/\lambda) = f(z/\lambda) | dz/dx |$ evaluated at $y = h^{-1}(x)$, if $| dz/dx |$ is just a scale factor, independent of $\lambda$, then the MLM estimate is $\hat{\lambda}_{ML} = {1}/{\bar{X}} = {1}/{\bar{h(Z)}}$ so that
\begin{equation}
\hat{\eta}^Z_{\alpha,ML} = h^{-1} \left( - \frac{\log( 1 - \alpha)}{n} \sum h(z_i) \right).
\end{equation}
The MLM, like the Bayesian method, is invariant under $h$ although, in general, the MLM quantile estimate will no longer be unbiased.

More generally, if $Z$ is a random variable with cdf $F(z) = 1 - e^{-\lambda h(z)}$ with $\lambda > 0$, $h(z) \geq 0$, and $h$ strictly increasing, or $F(z) = e^{-\lambda h(z)}$ with $h$ strictly decreasing, and the support is $[u, \infty)$ and h(u) = 0 (or $(u, \infty)$ and $\lim_{w \to u} h(w) = 0$) then a ZCE quantile estimate is constructed as $\hat{\eta}_\alpha^Z = h^{-1}( \Psi \sum h(z_i) )$ with $\Psi = (1-\alpha)^{-1/n} - 1$ or $\Psi = (\alpha)^{-1/n} - 1$, respectively.

In the first case, we can show that $X = -\log( 1 - F_1(Z) ) \sim Exp(\lambda)$ where $F_1(\cdot)$ is $F(\cdot)$ with $\lambda = 1$. In this case, $X = -\log( 1 - (1-e^{-h(z)}) ) = h(z)$ and $dX/dZ = h'(x)$ so by the usual change of variables
\begin{equation}
    f_X(x) = \frac{ f_Z(z) }{|dX/dZ|} = \frac{\lambda h'(x) e^{-\lambda h( h^{-1}(x))}}{h'(x)} = \lambda e^{-\lambda x}
\end{equation}
and similarly for the second case.  An example with numerous applications is the Rayleigh distribution, $f_Z(z) = \frac{z}{\sigma^2} e^{-z^2/2 \sigma^2}$,  where $h(Z) = Z^2$. Another example is the standard Pareto distribution, with tail-index $1/\lambda$,
\begin{equation}
    F(z) = 1 - \left(\frac{z}{u}\right)^{-\lambda} = 1 - e^{-\lambda \log(z/u)} \;\;\;\; z \geq u
\end{equation}
so that $h(Z) = \log(Z/u)$ and a ZCE quantile estimate is
\begin{equation}
    \hat{\eta}_\alpha^Z = u e^{ \Psi \sum \log(z_i/u) } \;\;\;\; \Psi = (1-\alpha)^{-1/n} - 1
\end{equation}
assuming $u$ is fixed. In the next section, we use this result to model the conditional tail of distributions that have asymptotic Paretian tails such as those in the Fréchet maximum domain of attraction.

It is worth noting that if $Z \sim F(Z)$ can be transformed to a uniform distribution, $U[0,1]$, then $-\log(F(Z)) \sim Exp(1)$ hence a ZCE quantile estimate can be formed.  This approach, suggested by (\citet{YuAlly2009}), can be used when the parameters of the cdf, $F$, are known such as if the scale parameter $\lambda = 1$. 

\section{ZCE Quantile Estimation for Paretian Tails}
\label{Sec3}

\subsection{Extreme Value Theory}
\label{Sec3.1}

The theory of extremes is a well-established branch of probability theory and statistical science (\citet{colesbook}, \citet{gumbelbook}). Consider the maximum, $M_n = \max\{ X_1, X_2, \cdots, X_n\}$, of i.i.d. random variables ({RVs}) with common distribution function (cdf) $X \sim G(x)$. If the cdf of $M_n \sim G^n(x)$, properly normalized, converges to a non-degenerate distribution, the Fisher-Tippet-Gnedenko (FTG) theorem states that it is in the generalized extreme value (GEV) family with tail-index $\xi$ (\citet{ekm_MEE_2003}) . This distribution has the Jenkinson-von Mises representation,
\beq
\label{GEVcdf}
H_{ \xi }(y) = \exp( - (1+ \xi y )^{-1/\xi} ),
\eeq
and we say $G$ is in the maximum domain of attraction of $H_{\xi}$ denoted as $G \in MDA(H_{\xi})$. Our interest lies with $\xi > 0, \; y \geq -1/\xi$, resulting in the standard Fr\'{e}chet distribution which is the limiting class for many underlying heavy-tailed distributions. By the convergence in types theorem, all normalizing sequences converge to the same limit so, more generally, we can state $G \in MDA(H_{\xi, \mu, \beta})$ with location $\mu$ and scale $\beta$.

The excess distribution, or conditional tail, is defined as $G_u(x) = \mathcal{P}( X-u \leq x | X > u)$ and the {Pickands-Balkema-de Haan} theorem states that there exists a positive-measurable function $\beta(u)$ such that
\beq
\lim_{u \to \infty} \sup_{x \geq 0} | G_u(x) - G_{\xi, 0, \beta(u)}(x) | = 0 
\eeq
if and only if $G \in MDA(H_\xi)$ where $G_{\xi, \mu, \beta}$ is the generalized Pareto distribution (GPD) with the same tail-index as the GEV (\citet{balkemahaan1974}, \citet{pickands1975}). 

Once again, more generally, if $G \in MDA(H_{\xi, \mu, \beta})$, the conditional tail distribution, for $x \geq u$, is
\beq
G_u(x) = P\left[ X < x | X > u \right] \approx 1-\left[ 1 + \frac{x-u}{\beta/\xi + u - \mu} \right]^{-1/\xi} \sim G_{\xi, u, \beta + \xi(u-\mu)}. 
\eeq
and a simplifying assumption (\citet{dej_pmd_ICASSP20}) is that the support of $M_n$ is $[0,\infty)$, or $\mu = \beta/\xi$, and the conditional tail reduces to a standard Pareto,
\beq 
G_{\xi, u}(x) = 1 - \left( \frac{x}{u} \right)^{-1/\xi} \;\; x \geq u, 
\eeq
with scale parameter $u$. In addition to reducing the parameter set, the Pareto RV can be transformed to an exponential distribution which is the basic assumption used for traditional methods of tail index, and large quantile, estimation (\citet{dekkershaan1989}, \citet{hill1975}, \citet{pickands1975}).

Given a high threshold, $u$, the point process of exceedances converges to a compound Poisson process and is commonly referred to as the 'Peaks over Threshold' (POT) model (\citet{Leadbetter1991}). Methods for threshold selection (\citet{northrop2017}) or marginalization of $u$ (\citet{dej_pmd_ICASSP21}) have been investigated, although often the threshold is set to include a fixed number of high exceedances (\citet{buishand1989}).

\subsection{Data Definition and Solution}

The underlying data consist of $\tilde{n}$ ''blocks" with the $i$'th block having length $m_i$ and, to be concrete, we assume each block is a year. The exceedances $\{x_1, \cdots, x_n\}$ are culled from the data by choosing the threshold $u$ equal to the $(n+1)$' first largest-order statistic.  This results in exceedances that are in the top $100 \times \frac{n}{\tilde{n}\bar{m}}$ percentile with $\bar{m}$ the average number of observations per year. In some fields, such as hydrology, the resulting samples are referred to as the partial duration series, or annual exceedance series when $n = \tilde{n}$. On average, there are $\frac{n}{\tilde{n}}$ observations per year and, if $\bar{m}$ is large enough, the POT model is justified.

Since a Pareto RV is isomorphic to an exponential RV, we denote the exceedances as $x_{1:n} = \{\ln(x_1/u), \cdots, \ln(x_n/u)\}$, which are i.i.d. and have distribution $F(x/\lambda) \sim$ Exp($\lambda$) with $\lambda = 1/\xi$. Furthermore, the number of exceedances are Poisson, Poi($\lambda_u$), so that our observations, or {\em training} data, $(n, x_{1:n})$, form a marked Poisson process with both parameters, $\lambda$ and $\lambda_u$, unknown. Given $N$ future years, we let $y_{1:N_u}$ denote the $N_u$ threshold exceedances of $u$ and $N_\alpha$ the number of those exceedances above $\hat{\eta}_\alpha$.  Typical levels for $\alpha$ range from 99\% to an extreme 99.99\%. As before, we wish to determine the unconditional distribution of future exceedances of $\hat{\eta}_\alpha$ or $\mathcal{P}(N_\alpha = k)$ for $k=0,1,...N_u$ and, in particular, $E[N_\alpha]$.

Given our $u$ threshold, $N_u$ is unknown but follows a Poisson process with rate parameter $\lambda_u$. The maximum likelihood estimate, for $E[N_u]$, would simply be $\hat{N_u} = \frac{n}{\tilde{n}} N$ based on a Poisson posterior with $\hat{\lambda_u} = \frac{n}{\tilde{n}}$. Given the partial duration series, with $n$ samples in $\tilde{n}$ years, the posterior, $\mathcal{P}(\lambda_u/n)$ is a $\Gamma(n+a, \tilde{n}+b)$ distribution, using a $\Gamma(a,b)$ conjugate prior, and the predictive distribution is therefore a negative binomial 
\beq
N_u \sim \text{NB}\left( n+a, \frac{\tilde{n}+b}{\tilde{n} + N + b} \right).
\eeq
Using a Jeffreys prior, with $a = 1/2, b = 0$, results in 
\beq
E[N_u] =  \frac{n}{\tilde{n}} N (1 + \frac{1}{2n} )
\eeq
and if $n = \tilde{n} >> 1$ then $E[N_u] \approx N$. 

Given the results from Section \ref{Sec2.1}, we can write the conditional expectation for the number of quantile estimate exceedances as
\begin{equation}
 \label{eq:exp_Na}
 E[ N_\alpha | N_u ] = \frac{N_u}{(\Psi + 1)^n}
\end{equation}
and, since we desire a ZCE quantile estimator with $\mathcal{P}( y > \hat{\eta}_{\alpha, Bayes}) = 1 - \alpha$, our Bayesian estimate requires
\beq
{\Psi} = \left( \frac{ \frac{n}{\tilde{n}} (1 + \frac{1}{2n} )}{1-\alpha}\right)^{1/n} - 1
\eeq
so that the unconditional estimate $E[ N_\alpha ] = 1$ and, as before, $\alpha$ is the {\em annual} quantile. We note that if $n = \tilde{n} >> 1$, then $\Psi \approx ({1-\alpha})^{-1/n} - 1$ as for the unconditional estimate.  If $n = \tilde{n}$ but small, then 
\beq
{\Psi} \approx \left(  \frac{ (1 + \frac{1}{2n} )}{1-\alpha}\right)^{1/n} - 1
\eeq
and $\hat{\eta}_{\alpha}$ is higher than the unconditional estimate to account for the added uncertainty in the number of $u$ exceedances, $N_u$. If If $n < \tilde{n}$ but large then 
\beq
{\Psi} \approx \left(  \frac{ n/\tilde{n}}{1-\alpha}\right)^{1/n} - 1
\eeq
and $\hat{\eta}_{\alpha}$ is lower than the unconditional estimate because the threshold $u$ is lower. In effect, we do not need as high of a conditional quantile to achieve the equivalent {\em annual}, unconditional, quantile. The ML estimate, would result in a similar value of $\Psi$ as for the unconditional distribution
\beq
{\Psi_{ML}} = \left( \frac{ \frac{n}{\tilde{n}} N }{1-\alpha}\right)^{1/n} - 1.
\eeq

The unconditional distribution for exceedance count is
\beq
P( N_\alpha = k ) = \sum_{N_u = k}^{N_m} BEG(k; n, N_u, \alpha) \times NB( N_u; n+\frac{1}{2}, \frac{\tilde{n}}{\tilde{n}+N}) 
\eeq
where the upper limit is the total number of test samples.  In practice, the summation can be truncated and simulation results show that using $N_u=E[N_u] = N$ in Equation \ref{beg_dist} does not affect the exceedance distribution significantly (See Figure \ref{fig:fig4}).

\section{Simulation Results}
\label{Sec4}

\begin{figure}[h]
	\centering
    \includegraphics[width=.65\textwidth]{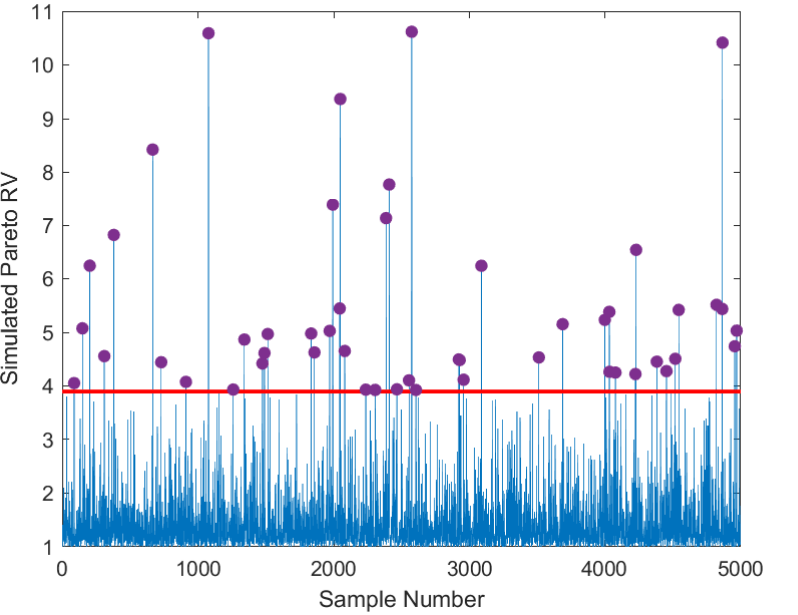}
    \caption{Simulated Pareto RV with $\xi = .3$, $n = \tilde{n} = 50$, $\bar{m}=100$}
    \label{fig:fig3}
\end{figure}

To illustrate the performance of the Bayesian and MLM quantile estimate, we began with a simulation of a standard Pareto RV with tail-index $\xi = .3$ an example of which is shown in Figure \ref{fig:fig3}.  We used $\tilde{n} = 50$ with $\bar{m} = 100$ samples per year.  The threshold $u$ was chosen to select the $n = 50$ extreme events highlighted in the Figure.  Given the underlying data are Pareto, we might expect the Bayesian method to perform well and, also, we could apply our results directly to all of the data if desired. We assumed $N$ = 100 years and we performed 10,000 runs to gather statistics. 

\begin{figure}[h]
	\centering
    \includegraphics[width=.65\textwidth]{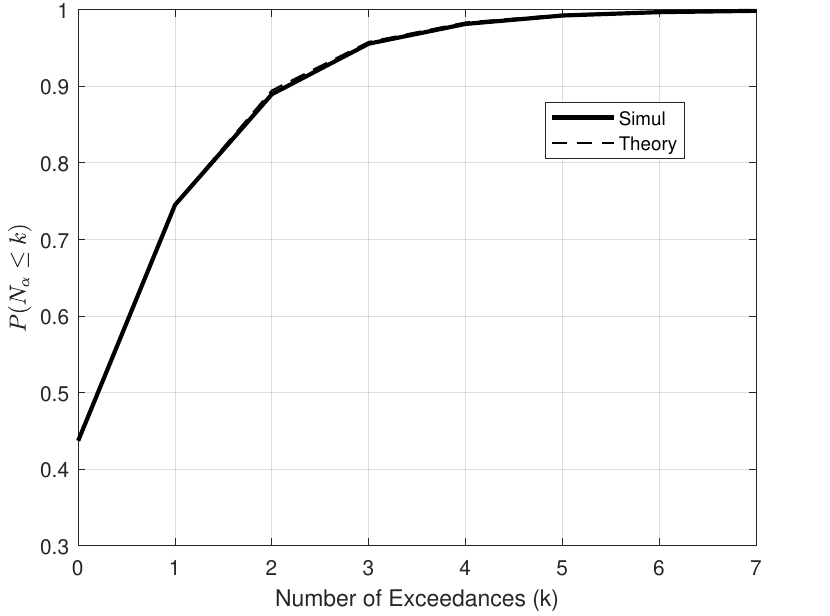}
    \caption{Theoretical $BEG(k; 50, 100, 99\%)$ and simulation frequency distribution for Bayesian quantile estimator.}
    \label{fig:fig4}
\end{figure}

Figure \ref{fig:fig4} shows the theoretical $BEG(k; 50, 100, 99\%)$ distribution along with the observed frequency distribution from the simulation using the Bayesian quantile estimator.  The results agree and, while the expected number of exceedances is one, there is a significant probability of more than one exceedance, which is roughly 25\%.  In some applications, where an exceedance implies a crititcal failure, this point may be moot, but the exceedance distribution is insightful for risk managers and policy makers to consider. In fact, a more desirable quantile estimator might be to solve for $\Psi$ in Equation \ref{beg_dist} such that $\mathcal{P}(N_\alpha = 0 )$ a fixed value (e.g., 50\%). 

\begin{figure}[h]
	\centering
    \includegraphics[width=1\textwidth]{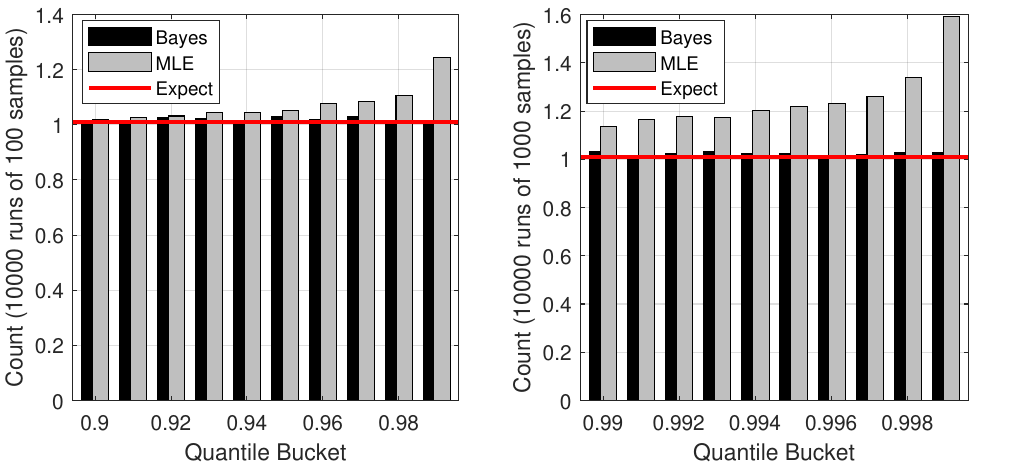}
    \caption{Normalized simulation exceedance count for 1\% prediction intervals with $n$ = 50 and $N = 100$ (left) and .1\% prediction intervals with $n$ = 50 and $N = 1000$ (right) }
    \label{fig:fig5}
\end{figure}

\FloatBarrier

Figure \ref{fig:fig5} (left) shows the simulation exceedance count divided by the number of simulation runs for the prediction intervals of 1\% from 90\% to 99\% with $N = 100$ for both the Bayes and the MLM quantile estimators. The Bayesian estimator performs as advertised with an average number of 1 sample per simulation run which compares the the MLM, which underestimates the quantile, in terms of coverage error, resulting in excessive exceedances.

Figure \ref{fig:fig5} (right) shows the more stressful case with prediction intervals of .1\% from 99\% to 99.9\% with $N = 1000$.  Once again, the Bayesian results in ZCE while the MLM performs rather poorly, particularly at the highest 99.9\% quantile. While such a stressful scenario of trying to estimate a 99.9\% quantile given only 50 samples may seem unlikely, we note that similar estimates have been required in response to natural disasters (\citet{dantzig1954}).

As expected, our method works well when the underlying distribution has a Paretian tail. To explore how the estimator performs when the underlying distribution has an asymptotic Paretian tail, we ran the simulation using a stable $\alpha_s$ symmetric (S$\alpha$S) distribution for different values of the characteristic exponent $\alpha_s$.  We do not necessarily advocate using our method if an S$\alpha$S distribution is appropriate and, in that context, estimating $\alpha_s$ and quantiles directly is advised (\citet{dumouchel1983}).  As in our first example, we used $\tilde{n} = 50$ with $\bar{m} = 100$ samples per year resulting in 5000 total samples. We used $N$ = 100 years with $\alpha = 99\%$ and we performed 10,000 runs.

\begin{figure}[h]
	\centering
    \includegraphics[width=1\textwidth]{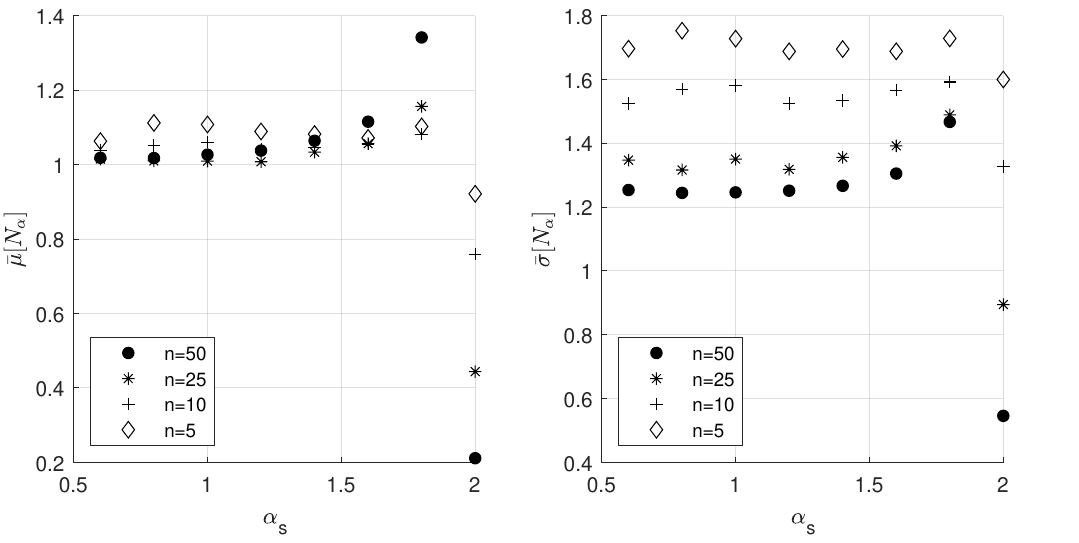}
    \caption{Simulation sample mean (left) and standard deviation (right) of the number of exceedances ($N_\alpha$) as a function of S$\alpha$S characteristic exponent $\alpha_s$ for different tail-samples ($n$) with $\tilde{n} \bar{m}$ = 5000, $N$ = 100, and $\alpha = 99\%$}
    \label{fig:fig6}
\end{figure}

As can be seen in Figure \ref{fig:fig6}, the performance of our quantile estimate is sensitive to the number of tail-samples chosen ($n$) as $\alpha_s \to 2$. This is not surprising, as the inflection point, beyond which a Paritian tail can be deemed appropriate, increases as $\alpha_s \to 2$ (\citet{hfofack1999}). This results in a need to reduce the number of tail-samples (increase threshold $u$) to $\approx$ .1\% of total samples although this comes with the cost of higher variability unless the total number of samples ($\tilde{n} \bar{m}$) can be increased.

For values of $\alpha_s < 1.5$, the method performs well for all values of $n$ with $n = 50$ or 1\% of the total samples resulting in the best performance with the lowest variability. Since $\alpha_s < 1.5$ implies a heavy-tail with most of the distribution having power-law decay, there is less bias introduced to the quantile estimate.  When $\alpha_s = 2$, a normal distribution not in the Fréchet domain, the sensitivity of the quantile to outliers is evident as is documented in the literature (\citet{weron2001}).

\begin{table}[h]
\centering
\caption{Simulation statistics for the tail-index, $\xi$, and the number of exceedances, $N_\alpha$, for various distributions and for different tail-samples ($n$) with $\tilde{n} \bar{m}$ = 5000, $N$ = 100, and $\alpha = 99\%$}
\vspace{.1in}
\label{table:t1}
\begin{tabular}{@{}lcccccccc@{}}
\hline
\multicolumn{1}{c}{} &
\multicolumn{4}{c}{Exp($\mu = 1$)} &
\multicolumn{4}{c}{LogNormal($\mu = $0, $\sigma =$1)} \\
\cline{1-9}
 &
\multicolumn{1}{c}{n = 5} &
\multicolumn{1}{c}{n = 10} &
\multicolumn{1}{c}{n = 25}&
\multicolumn{1}{c}{n = 50}&
\multicolumn{1}{c}{n = 5} &
\multicolumn{1}{c}{n = 10} &
\multicolumn{1}{c}{n = 25}&
\multicolumn{1}{c}{n = 50} \\
\hline
$\bar{\mu}[\xi]$ & .13 & .14 & .16 & .18 & .28 & .29 & .32 & .34 \\
$\bar{\sigma}[\xi]$  & .05 & .04  & .03   & .02 & .12 & .09  & .06   & .04 \\
$\bar{\mu}[N_\alpha]$   & .95 & .78  & .48   & .25 & .98 & .90  & .71   & .53 \\
$\bar{\sigma}[N_\alpha]$   & 1.67 & 1.37 & .93  & .60 & 1.60 & 1.45 & 1.12  & .88 \\
$\bar{\mathcal{P}}(N_\alpha > 1)$   & .21 & .18   & .10   & .04 & .23 & .22   & .17   & .11  \\
\hline
\end{tabular}
\begin{tabular}{@{}lcccccccc@{}}
\hline
\multicolumn{1}{c}{} &
\multicolumn{4}{c}{StdPar($\xi = .1$)} &
\multicolumn{4}{c}{GEV($\xi = .5, \beta = 1, \mu = 0$)} \\
\cline{1-9}
 &
\multicolumn{1}{c}{n = 5} &
\multicolumn{1}{c}{n = 10} &
\multicolumn{1}{c}{n = 25}&
\multicolumn{1}{c}{n = 50}&
\multicolumn{1}{c}{n = 5} &
\multicolumn{1}{c}{n = 10} &
\multicolumn{1}{c}{n = 25}&
\multicolumn{1}{c}{n = 50} \\
\hline
$\bar{\mu}[\xi]$ & .10 & .10 & .10 & .10 & .52 & .51 & .53 & .54 \\
$\bar{\sigma}[\xi]$  & .04 & .03  & .02   & .01 & .23 & .16  & .10   & .07 \\
$\bar{\mu}[N_\alpha]$   & 1.08 & 1.04  & 1.03   & 1.0 & 1.05 & 1.03  & .94   & .88 \\
$\bar{\sigma}[N_\alpha]$   & 1.70 & 1.53 & 1.37  & 1.22 & 1.70 & 1.50 & 1.30  & 1.14 \\
$\bar{\mathcal{P}}(N_\alpha > 1)$   & .26 & .25   & .26   & .26 & .25 & .25   & .23   & .23  \\
\hline
\end{tabular}
\begin{tabular}{@{}lcccccccc@{}}
\hline
\multicolumn{1}{c}{} &
\multicolumn{4}{c}{StudentT($\nu = 2$)} &
\multicolumn{4}{c}{StudentT($\nu = 10$)} \\
\cline{1-9}
 &
\multicolumn{1}{c}{n = 5} &
\multicolumn{1}{c}{n = 10} &
\multicolumn{1}{c}{n = 25}&
\multicolumn{1}{c}{n = 50}&
\multicolumn{1}{c}{n = 5} &
\multicolumn{1}{c}{n = 10} &
\multicolumn{1}{c}{n = 25}&
\multicolumn{1}{c}{n = 50} \\
\hline
$\bar{\mu}[\xi]$ & .50 & .50 & .50 & .51 & .14 & .15 & .17 & .18 \\
$\bar{\sigma}[\xi]$  & .23 & .16  & .10   & .07 & .06 & .04  & .03   & .02 \\
$\bar{\mu}[N_\alpha]$   & 1.08 & 1.04  & 1.0   & .95 & 1.02 & .86  & .64   & .40 \\
$\bar{\sigma}[N_\alpha]$   & 1.71 & 1.57 & 1.34  & 1.19 & 1.69 & 1.39 & 1.07  & .73 \\
$\bar{\mathcal{P}}(N_\alpha > 1)$   & .25 & .25   & .25   & .24 & .24 & .20   & .14   & .08  \\
\hline
\end{tabular}

\end{table}

\FloatBarrier

To further investigate the performance of the Bayesian quantile estimator, we applied it to different underlying distributions with statistics reported in Table \ref{table:t1}. Similar to the results when $\alpha_s = 2$, the estimator is too conservative for thinner, exponential-tailed distribution such as the exponential or lognormal.  These distributions, which belong to the Gumbel maximum domain of attraction, require less tail-samples, about .1\% of total samples, to avoid outlier-bias issues.  This may not be that surprising as we are fitting a power-law tail to exponential decay or, effectively, using the wrong tool for the job.  

The method works well for the standard Pareto even when the tail index is closer to 0 (i.e., $\xi = .1$) and also for the GEV, or Fréchet, distribution.  We specifically chose $\mu = 0 \neq \beta/\xi$ so that our condition to simplify the conditional tail to a standard Pareto is violated (See Section \ref{Sec3.1}).  The variance-bias trade-off is seen as the average number of exceedences declines to .88 for 1\% sample selection ($n = 50$) while the variance in the number of exceedences also declines. Lastly, we show the simulation results for two Student T distributions. A heavy-tailed version with degrees of freedom, $\nu = 2$, and lighter tail with $\nu = 10$. 

In this research, we have derived a Bayesian quantile estimator and the distribution of exceedances that can be applied to distributions whose tail exhibits power-law decay.  The theoretical performance exceeds the ML method and simulation results agree. Our method can be applied to a variety of underlying distributions, although care in threshold selection is required for thinner tailed distributions that are outside the Fréchet domain of attraction.  Given financial data, such as stock market returns, or climate data, such as precipitation, appear to be in the Fréchet domain with tail-index values in the .25 to .5 range, we expect our method to be a valuable alternative to traditional quantile estimation techniques.

\bibliographystyle{unsrtnat}
\bibliography{DEJRefs}  

\end{document}